\documentclass[usenatbib]{mn2e} 
\usepackage{longtable} 
\usepackage{pdflscape} 
\usepackage{amssymb, textcomp, verbatim, graphicx, times, url,subfigure} 
\usepackage{amsmath} 
\usepackage{xcolor}
\usepackage{soul}
\usepackage[compact]{titlesec}
\titlespacing{\section}{0pt}{2ex}{1.5ex}
\titlespacing{\subsection}{0pt}{2ex}{1.5ex}
\usepackage[bottom]{footmisc}

\makeatletter
\def\@makefnmark{\hbox{\@textsuperscript{\normalfont\@thefnmark}}}

\def\aj{AJ}
\def\apj{ApJ}
\def\apjl{ApJ}
\def\aa{A\&A}
\def\mnras{MNRAS}
\def\nat{Nature}

\title{A GMRT 150 MHz Search for Variables and Transients in Stripe 82}
\author[Hajela et al.]{A. Hajela$^1$, K.~P. Mooley$^{2,3,9}$, H.~T. Intema$^4$ \& D.~A. Frail$^2$ \\
  $^{1}$ Department of Physics and Astronomy, Northwestern University, Evanston, IL 60208. ahajela@u.northwestern.edu\\
  $^{2}$ National Radio Astronomy Observatory, P.O. Box O, Socorro, NM 87801.\\
  $^{3}$ Cahill Center for Astronomy, MC 249-17, California Institute of Technology, Pasadena, CA 91125, USA.\\
  $^{4}$ Leiden Observatory, Leiden University, Niels Bohrweg 2, NL-2333CA, Leiden, The Netherlands\\
  $^{9}$ Jansky Fellow
  }
  
\begin{document}
\maketitle

\begin{abstract}
We have carried out a dedicated transient survey of 300 deg$^2$ of the SDSS Stripe 82 region using the Giant Meterwavelength Radio Telescope (GMRT) at 150 MHz.
Our multi-epoch observations, together with the TGSS survey, allow us to probe variability and transient activity on four different timescales, beginning with 4 hours, and up to 4 years.
Data calibration, RFI flagging, source finding and transient search were carried out in a semi-automated pipeline incorporating the SPAM recipe.
This has enabled us to produce superior-quality images and carry out reliable transient search over the entire survey region in under 48 hours post-observation.
Among the few thousand unique point sources found in our 5$\sigma$ single-epoch catalogs (flux density thresholds of about 24 mJy, 20 mJy, 16 mJy and 18 mJy on the respective timescales), we find $<$0.08\%, 0.01\%, $<$0.06\% and 0.05\% to be variable (beyond a significance of 4$\sigma$ and fractional variability of 30\%) on timescales of 4 hours, 1 day, 1 month and 4 years respectively.
This is substantially lower than that in the GHz sky, where $\sim$1\% of the persistent point sources are found to be variable.
Although our survey was designed to probe a superior part of the transient phase space, our transient sources did not yield any significant candidates. 
The transient (preferentially extragalactic) rate at 150 MHz is therefore $<$0.005 on timescales of 1 month and 4 years, and $<$0.002 on timescales of 1 day and 4 hours, beyond 7$\sigma$ detection threshold.
We put these results in the perspective with the previous studies and give recommendations for future low-frequency transient surveys.

\end{abstract}

\begin{keywords}
      catalogues --
      galaxies: active --
      stars : activity --
      radio continuum: galaxies --
      surveys
\end{keywords}

\section{INTRODUCTION}\label{sec:intro}


\noindent Our understanding of the dynamic radio sky on timescales $>$1s has relied heavily on the radio follow up of transients discovered through synoptic surveys at optical, X-ray, or gamma-ray wavelengths. 
However, a significant fraction of transients, such as the ones residing in dust-obscured environments, those powered by coherent emission processes, and unbeamed phenomena, are missed by these synoptic surveys.
Blind radio searches have the exceptional ability to access this population of transients, thus giving an unbiased rate of these events. \\
\indent There has been significant progress made with blind searches at GHz frequencies over the past few years. Since the transient rates are low \cite[e.g.][]{frail2012}, these searches have highlighted the use of widefield observations together with near-real-time data processing and extensive follow up observations in order to maximize the transient yield and identification \citep{mooley2016}. 
Only a few percent of the persistent radio sources are found to be variable, with AGN dominating this sample \citep[e.g.][]{frail1994,carilli2003,deVries2004,croft2010,thyagarajan2011,bannister2011,ofek2011,williams2013,bell2015,mooley2016,hancock2016}.  
Widefield surveys have led to the discovery of several AGN showing renewed jet activity on timescales of $\sim$40,000 years, stellar explosions, a tidal disruption event, and flares from Galactic sources \citep{gal-yam2006,thyagarajan2011,bannister2011,mooley2016}.
Radio transient surveys such as the VLA Sky Survey (Lacy et al. in prep) with the Karl G. Jansky Very Large Array (VLA), the ThunderKAT program on the MeerKAT telescope \citep{fender2017} and the ASKAP Survey for Variables and Slow Transients \cite[VAST;][]{murphy2013} program, will substantially increase the number of radio transients (at GHz frequencies) in the coming years.


On the other hand, blind searches for transients at MHz frequencies have had limited success.
With modest sensitivities, the vast majority of these surveys\footnote{A fairly complete compilation of radio transient surveys carried out till date can be found 
at \url{http://www.tauceti.caltech.edu/kunal/radio-transient-surveys/index.html}.} have probed mainly the Jansky-level population, and the transient yield has been low.
The majority of the transients that were found have ambiguous or unknown classification due to the searches being carried out in archival data and untimely follow-up observations.

Nevertheless, the transients discovered thus far assure a rich phase space of the dynamic MHz sky.
\cite{hyman2005,hyman2007,hyman2009} discovered three "Galactic Center Radio Transients (GCRTs)", with peak flux densities ranging from of tens to thousands of mJy, among which one was a flaring X-ray binary and two transients were of unknown origin \citep[but one likely a coherent emitter;][]{ray2007,polisensky2016}. 
\cite{jaeger2012} reported a 2.1 mJy transient in the SWIRE Deep Field 1046+59 at 340 MHz with the VLA, with no known counterparts.
Another transient, possibly Galactic in origin and lasting for $<$10 min with a peak flux density of about 20 Jy, was discovered in $\sim$400 hours of LOFAR 30 MHz data towards the North Celestial Pole at 60 MHz \citep{stewart2016}.
\cite{obenberger2014} discovered two transients at 30 MHz, having peak flux densities of about 3 kJy, and lasting for 75--100 seconds with evidence for polarization or dispersion. 
\cite{murphy2017} recently found a transient, having a peak flux density of 180 mJy and timescale between 1--3 years, while comparing the TGSS-ADR \citep{intema2017} and GLEAM \citep{hurley-walker2017} catalogs.

The MHz transient sky is expected to be different from the GHz sky.
On timescales of $>$1 s, the GHz sky is illuminated primarily by (incoherent) synchrotron-driven transients arising from astrophysical shocks, such as supernovae, gamma-ray bursts, tidal disruption events, AGN, X-ray binaries, etc., and from astrophysical plasma accelerated in stellar magnetic fields observed in the form of stellar flares, magnetar flares, etc \citep[e.g.][]{mooley2016}.
Being brightness-temperature limited, these transients evolve on timescales of days--months (extragalactic; more luminous) or hours--weeks (Galactic; less luminous), as noted by \cite{pietka2015}.
Most classes of incoherent synchrotron transients are self-absorbed at MHz frequencies at early times, pushing these events to much longer timescales of years to decades and lower peak flux densities compared to GHz frequencies.
Consequently, their rates are lower, and they are harder to identify in transient surveys (\citealt{metzger2015}).
On the other hand, transients powered by coherent emission (such as pulsars and brown dwarfs) may be more abundant at MHz frequencies.

Likewise, we expect the variable MHz sky to be different as well. 
Rather than the substantial intrinsic variability observed in the GHz sky, variability at MHz frequencies  will be dominated by refractive interstellar scintillation \cite[e.g.][]{rickett1986}. 
Interplanetary scintillation \citep{clarke1964,morgan2018}, caused due to local density fluctuations in the ionised medium in the ecliptic plane, will dominate the extrinsic variability close to the ecliptic. 

Given the yield of transients at $\sim$Jansky flux densities in the low-frequency sky, one would expect a multifold increase in the yield by probing deeper, at milliJanky flux densities. 
Motivated by this, and the need for systematic exploration of the mJy-level dynamic sub-GHz sky, we have carried out a dedicated survey over 220 deg$^2$ of the SDSS Stripe 82 region with the GMRT at 150 MHz.
GMRT offers both good sensitivity and $\sim$arcsec localization; the latter is essential for associating radio variable/transient sources with their optical counterparts.
The choice of our survey region is motivated by the presence of the abundance of deep multiwavelength archival data in Stripe 82, which aids our search for the progenitors/host galaxies of transients. 
Using the dataset, we are able to probe timescales between $\sim$hours and $\sim$1 month.
The observing frequency of 150 MHz allows us to take advantage of the existing TGSS survey and extend our transient search to a timescale of $\sim$4 years.
In \S\ref{sec:obs} we describe the observations, the calibration and source cataloging procedures.
In \S\ref{sec:var_search} and \S\ref{sec:trans_search} we detail the variability and transient search.
The summary and discussion are given in \S\ref{sec:summary}.

\section{OBSERVATIONS AND DATA PROCESSING}\label{sec:obs}
\subsection{Observations}\label{sec:obs_obs}
\noindent Stripe 82 is an equatorial strip on the sky, spanning 2.5 degrees in declination between $\pm$1.25 degrees, and 109 degrees in right ascension between $-$50 degrees and +59 degrees. Since the half-power beamwidth (HPBW) of GMRT at 150 MHz is 186 arcmin, we were able to cover the declination range of Stripe 82 in a single pointing. In right ascension, the pointings were spaced by HPBW/2 to get a fairly uniform sensitivity across Stripe 82. 

We observed two regions, R1 and R2, in November--December 2014 and June--September 2015 under project codes 27\_032 and 28\_082 respectively. Twenty seven pointings centred on declination of 0 degrees and spanning 0--40 degrees in right ascension were used for region R1. Thirty pointings centred on declination of 0 degrees and spanning 310--355 degrees in right ascension were used for region R2. Data was recorded in full polarization mode every 8 seconds, in 256 frequency channels across 16 MHz of bandwidth (140--156 MHz). We observed each region in two epochs, 1 month apart, with each epoch being split over two observing sessions usually spread over two consecutive days. In a single session, typically 15--30 pointings (covering an area of 50--100 deg$^2$), with each pointing observed for 20--40 minutes split over 2 scans (each scan was 10--20 minutes long) spaced out in time (about 4 hours) to improve the UV-coverage. The flux calibrator, 3C48, was observed in the middle and beginning/end of each session. Due to the presence of in-beam calibrators and the use of the SPAM recipe for direction-dependent calibration \citep{intema2009}, no phase calibration scans were obtained. An overview of all GMRT observations used for the variability and transient search is given in Table~\ref{tab:gmrt_log}.

\subsection{RFI Flagging, Calibration and Imaging using the SPAM recipe}\label{sec:obs_proc}
\noindent After each observation, the data were downloaded from the GMRT archive within 12 hours onto the computer cluster at the NRAO in Socorro, and processed with a fully automated pipeline based on the SPAM recipe \citep{intema2009,intema2017}. The pipeline incorporates direction-dependent calibration and modeling of ionospheric effects, generally yielding high-quality images. In brief, the pipeline consists of two parts: a pre-processing part that converts the raw data from individual observing sessions into pre-calibrated visibility data sets for all observed pointings, and a main pipeline part that converts each pre-calibrated visibility data set per pointing into a Stokes I continuum image. Both parts run as independent processes on the multi-node, multi-core compute cluster, allowing for parallel processing of many observations and pointings. A detailed description of the processing pipeline is given in \citet{intema2017}. With this pipeline, we were able to calibrate and image each GMRT observation within 10 hours after retrieval.

In addition to imaging each pointing per observing run, we also imaged each pointing for every scan (typically two scans per observing run; see \S\ref{sec:obs_obs}) and every epoch (E1/E2; combining the visibility data from the observing runs on consecutive days). 

\begin{table}
\centering
\small
\caption{GMRT Observing Log}
\label{tab:gmrt_log}
\begin{tabular}{lcccc}
\hline\hline 
No. & Date        & Region/Epoch & LST     & RMS$^a$ \\
    & (UT)        &              & (h)     & (mJy/beam)\\
\hline
\multicolumn{5}{c}{Archival Data: TGSS}\\
\hline
0   & 2010 Dec 15$^b$ & R1\&2E0      & -- & $\sim 3.5$ \\
\hline
\multicolumn{5}{c}{G1STS Observations}\\
\hline
1   & 2014 Nov 10 & R1E1a        & 19--06  & 3.8\\ 
2   & 2014 Nov 11 & R1E1b        & 19--06  & 4.1\\ 
3   & 2014 Dec 27 & R1E2a        & 16--01  & 4.8\\ 
4   & 2014 Dec 28 & R1E2b        & 17--01  & 6.6\\ 
5   & 2015 Jun 29 & R2E1a        & 22--09  & 2.8\\ 
6   & 2015 Jun 30 & R2E1b        & 23--09  & 2.6\\ 
7   & 2015 Aug 31 & R2E2a        & 20--05  & 2.5\\ 
8   & 2015 Sep 02 & R2E2b        & 20--05  & 2.4\\ 
\hline
\multicolumn{5}{p{2.8in}}{$^a$RMS refers to the median single-pointing RMS noise achieved during the given observing run.}\\
\multicolumn{5}{p{2.8in}}{$^b$This is the median epoch of TGSS survey. The TGSS observations were taken over two years from April 2010 to March 2012.}
\end{tabular}
\end{table}

\begin{figure}
\includegraphics[width=3in]{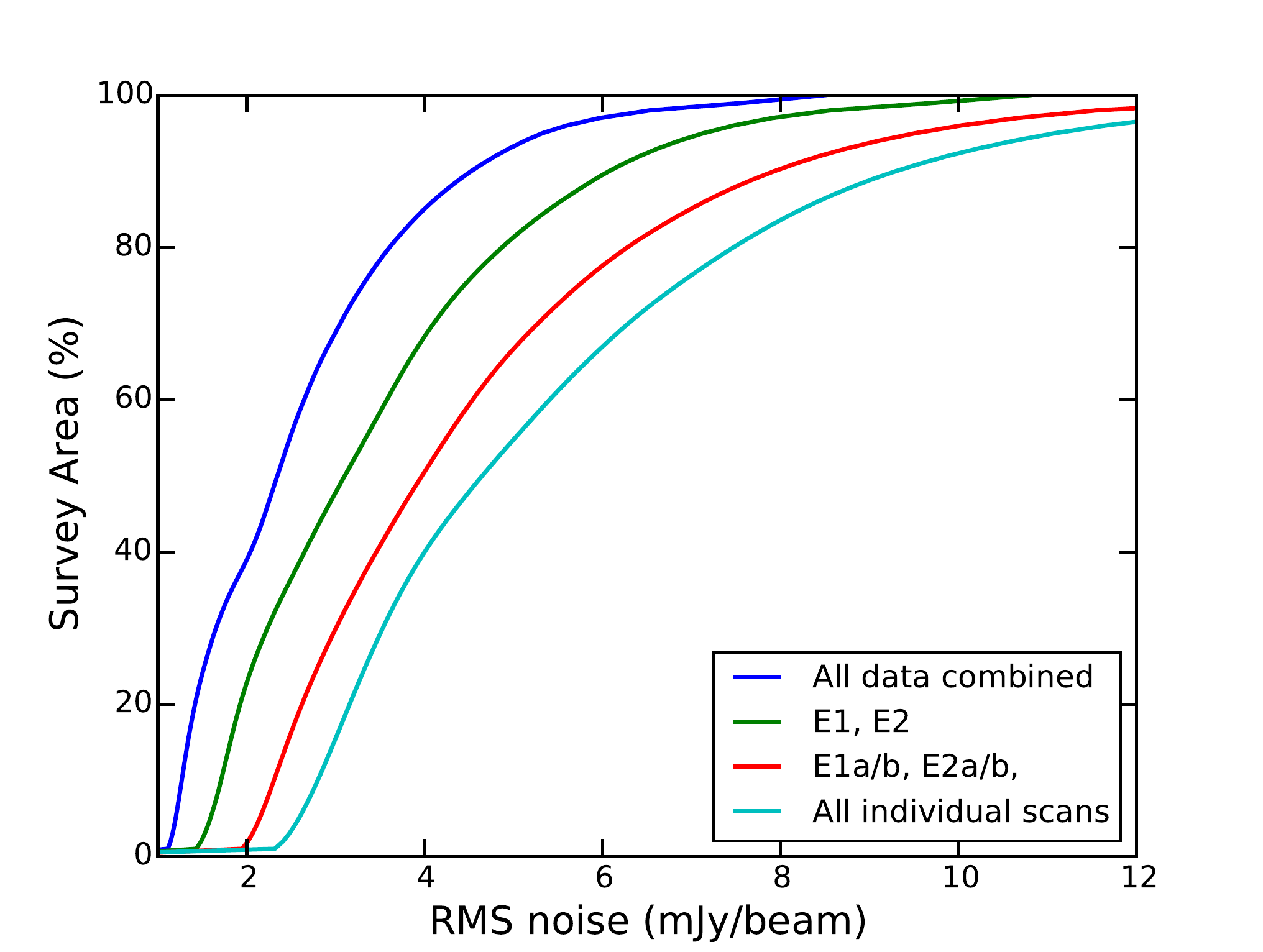}
\caption{Cumulative plot of the RMS noise for each timescale probed by the GMRT data. See \S\ref{sec:obs_obs}and Table~\ref{tab:gmrt_log} for details.}
\label{fig:rms_cum}
\end{figure}

\subsection{Image Mosaicing and Source Cataloging}\label{sec:obs_catalogs}
\noindent Once the single-pointing images were produced by the SPAM pipeline, we combined them into mosaics using the AIPS task FLATN. 
The RMS noise of the image mosaics generated for each scan, each observing run, each epoch and all data combined, are shown in Figure~\ref{fig:rms_cum}, and the median values for each observing run are reported in Table~\ref{tab:gmrt_log}.

We used PyBDSF\footnote{\cite{PythonBDSF}}, 
a Python module, to decompose images for every observing run, the corresponding scans and the epochs into sources and generate a 5$\sigma$ catalog. We used \texttt{process\_image} task of PyBDSF to process and find sources above a user-defined threshold in each individual image. \texttt{process\_image} offers a user-defined parameter, \texttt{rms\_box}, which was used to calculate the mean and the rms of the image using two inputs, the first fixed the rms-box size to calculate the mean and the rms and the second input fixed the step-size by which the box moved across the image. For this work, we used an rms box which was 20 times the size of the synthesized beam of the image \citep{hancock2012,mooley2013} and moved it by 10 pixels (i.e. the step-size) for the next measurement.
We used the module-default values for \texttt{thresh\_pix} = 5.0 and \texttt{thresh\_isl} = 3.0. The combination of these two parameters set the threshold for source detection in the images. \texttt{thresh\_isl} defined the threshold to select the regions or islands to which Gaussian is fitted and \texttt{thresh\_pix} defines the threshold for individual pixels to be included in that island. We wrote down all the detected sources and their properties in a catalog using \texttt{write\_catalog} task of PyBDSF.

The $\sim$300 deg$^2$ co-added image mosaics and the corresponding 5$\sigma$ source catalog containing 12,703 sources above 10.5 mJy is available via the Caltech Stripe 82 Portal\footnote{\url{http://www.tauceti.caltech.edu/stripe82/}}.

\subsection{Archival Data}\label{sec:obs_archival}
\noindent The Stripe 82 region is also covered by the 150~MHz GMRT sky survey TGSS\footnote{Details of the Alternative Data Release (TGSS-ADR) can be found in \cite{intema2017} and at \url{http://tgssadr.strw.leidenuniv.nl/}} with a very similar sensitivity ($\sim 3.5$~mJy/beam). The TGSS observations were performed over 2 years, from April 2010 to March 2012 with a median epoch of about 2010~Dec~15. We have used the publicly available data products from the TGSS-ADR to construct a 5$\sigma$ catalog of the same area in Stripe 82, which provides an extra epoch for our transient search (on $\sim$4 yr timescale). 

\section{VARIABILITY SEARCH}\label{sec:var_search}
\noindent From our GMRT observations of Stripe 82 alone, we can probe (via ``two-epoch" comparisons) variability on three timescales: 4 hours, 1 day and 1 month. As alluded to in \S\ref{sec:obs}, each of the eight observations listed in Table~\ref{tab:gmrt_log} was carried out using two scans separated by approximately four hours. Hence, in order to study the variability on this four hour timescale, we compared the 5-sigma source catalogs of the two scans\footnote{We excluded E1b from our analysis due to missing data and presence of substantial RFI.}. 
To study variability on a timescale of 1 day, we compared observation E1a with E1b, and observation E2a with E2b (cf. Table~\ref{tab:gmrt_log}). 
For the 1 month timescale, we compared E1 and E2 (obtained by combining E1a+E1b and E2a+E2b respectively, for regions R1 and R2; see \S\ref{sec:obs_proc}).
For the 4 year timescale, we compared our full combined dataset (all eight observations listed in Table~\ref{tab:gmrt_log} combined into a single deep mosaic) with the TGSS ADR1.
It should be noted that if a source is found to be variable between two epochs, its variability timescale is generally smaller than the separation between the two epochs and larger than the duration of each of the two epochs. For example, when comparing individual scans of each observation, we are probing a timescale of $<$4 hours (and $\gtrsim$30 min). 
A variable source will be unresolved at our angular resolution of $\sim 19 \arcsec$, unless that source is very nearby ($\ll$ 1 pc) and expanding extremely rapidly (superluminal motion).
Therefore, in order to shortlist point-like (unresolved) sources, and to avoid potential false sources/imaging artifacts, we applied the constraints listed below to the 5$\sigma$ catalogs: 


\begin{itemize}
    \item{\bf{Search area bounds}.} Due to very low sensitivity beyond $\sim$1.75 degrees from the GMRT 150 MHz beam center, the edges of our image mosaics of regions R1 and R2 are noisy. Hence we retained only those sources satisfying -1.75 deg $<$ Dec $<$ 1.75 deg, -1.25 deg $<$ RA $<$ 41.25 deg and 308.75 deg $<$ RA $<$ 356.25 deg.
	\item {\bf Flux density ratio.} Following \cite{mooley2016} and \cite{frail2018}, we keep sources having S/P $< 1.5$ (SNR$<$15) and S/P $<$ 1.1 (SNR$\geq$15), where $S$ is the total flux density and $P$ is peak flux density of the source.
    \item {\bf Source size.} We retained sources having\newline \texttt{BMAJ}/1.5$<$MAJ$<$1.5$\times$\texttt{BMAJ} and \texttt{BMIN}/1.5$<$MIN$<$1.5$\times$\texttt{BMIN}, where \texttt{BMAJ} and \texttt{BMIN} are the major and minor axis of the synthesized beam and MAJ, MIN are the major and minor axis of the Gaussian fitted by PyBDSF. We further imposed MAJ $>$ 1.1$\times$\texttt{BMAJ}, MIN $>$ 1.1$\times$\texttt{BMIN} for sources detected at a high significance (SNR $\geq$ 15) \citep[e.g.][]{mooley2016}.
	\item {\bf Proximity to bright sources.} To avoid any potential imaging artifacts around bright sources, we removed fainter sources (sources with total flux density $\ll$ 500 mJy) lying within $3$ arcmin of all $>$ 500 mJy sources.
\end{itemize}


\begin{figure}
\includegraphics[width=9cm,keepaspectratio]{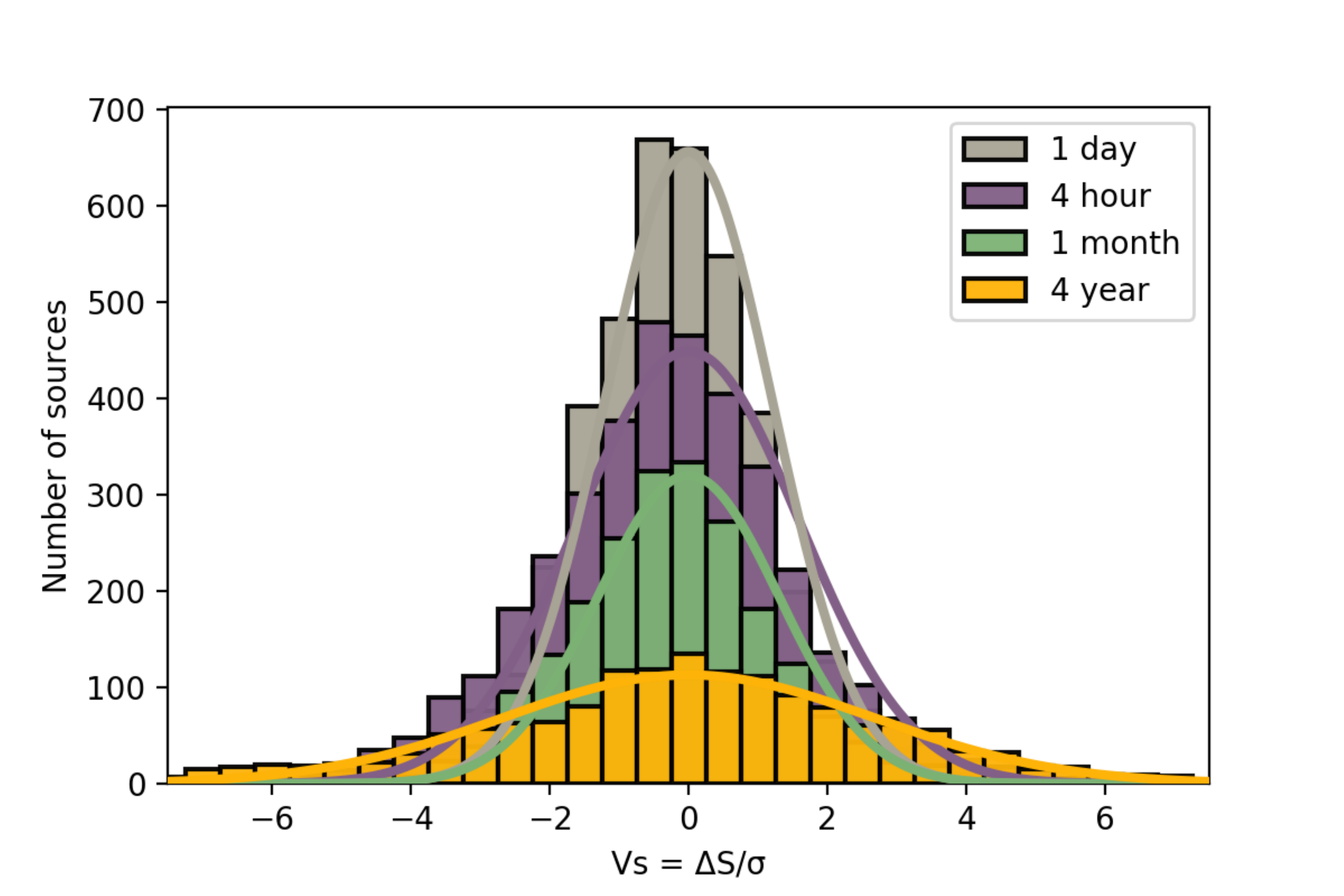}
\caption{The histograms of variability statistic $V_s$ corresponding to all timelines. $V_s$ is calculated after applying all the constraints to the single-epoch catalogs. Histograms are fit by the Gaussians of same color. Standard deviations, {\tt std}, of the fitted Gaussians for 4 hour timescale: 1.6, for 1 day timescale: 1.2, for 1 month timescale: 1.3 and for 4 year timescale: 2.7} 
\label{fig:Vs_hist1}
\end{figure}

\begin{figure*}
\centering
	\includegraphics[width=8cm,keepaspectratio]{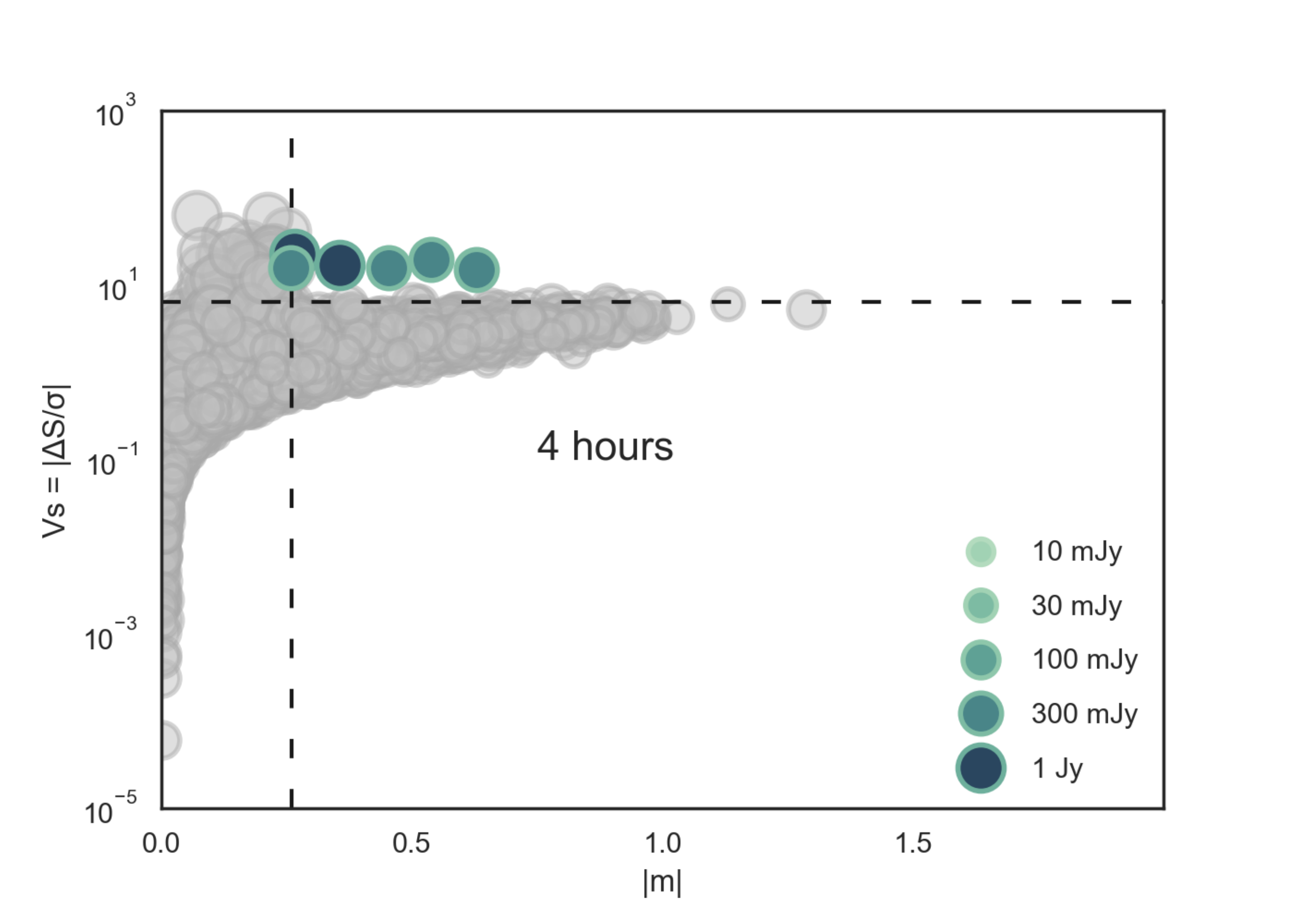}
	\includegraphics[width=8cm,keepaspectratio]{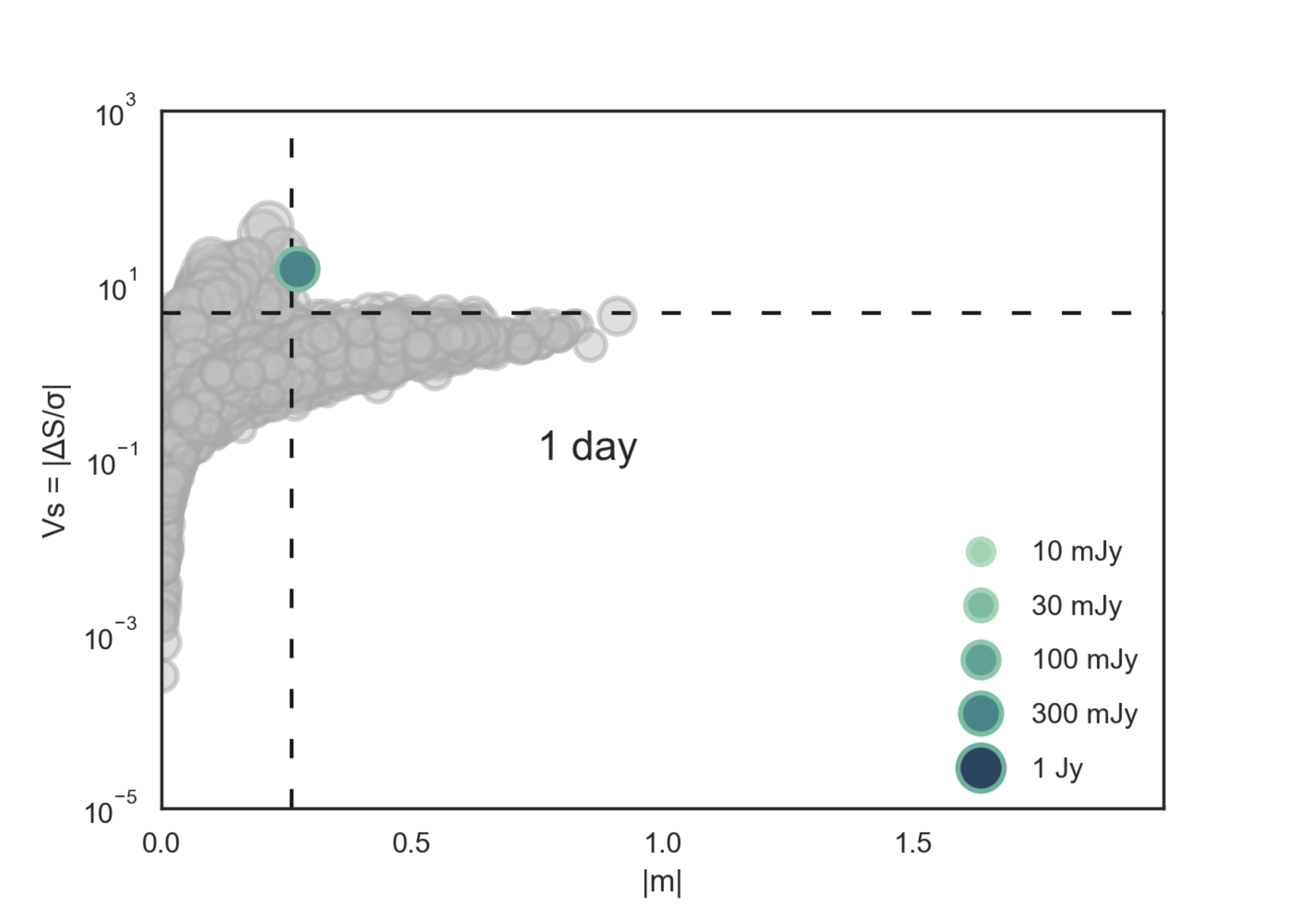}
	\includegraphics[width=8cm,keepaspectratio]{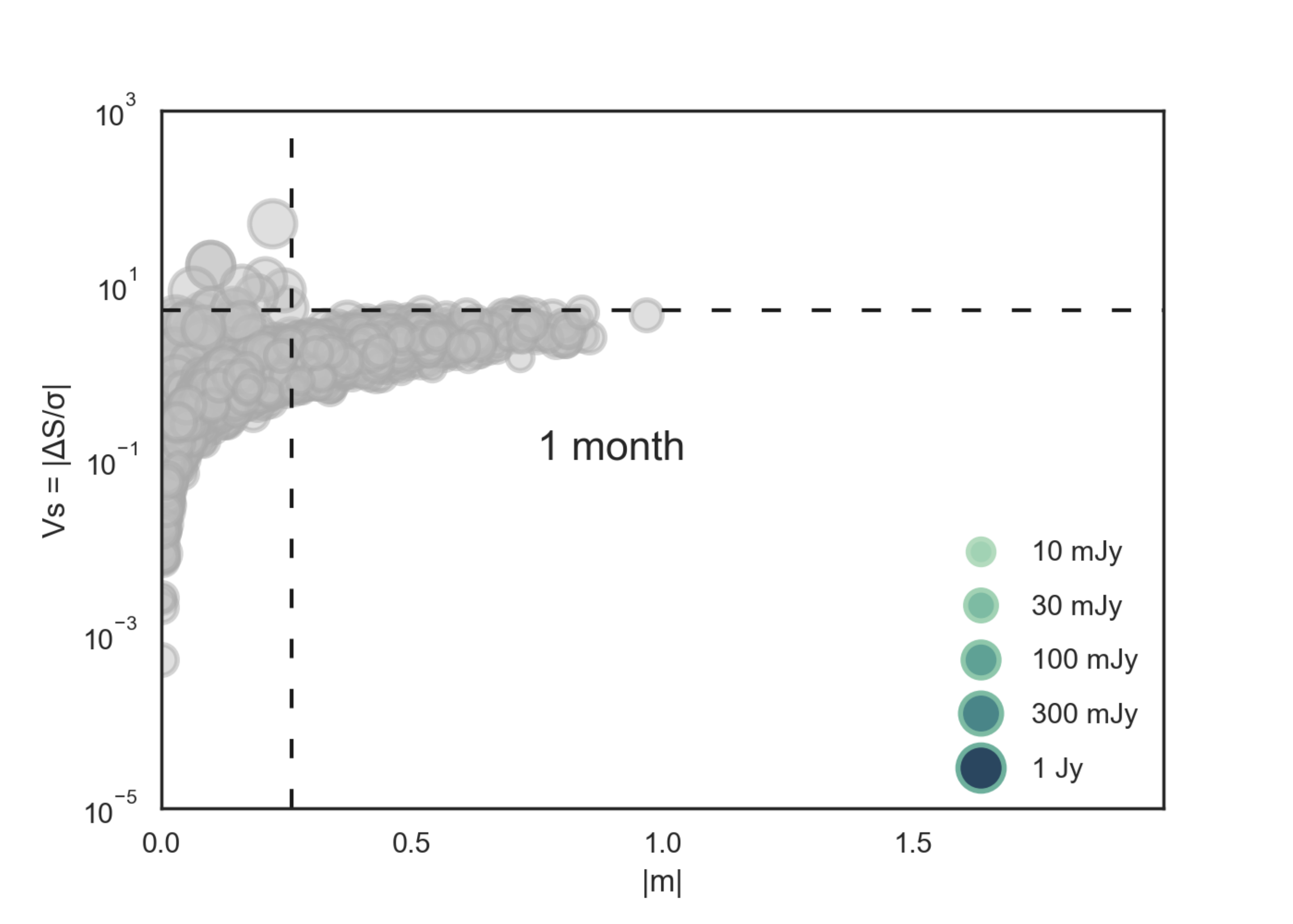}
    \includegraphics[width=8cm,keepaspectratio]{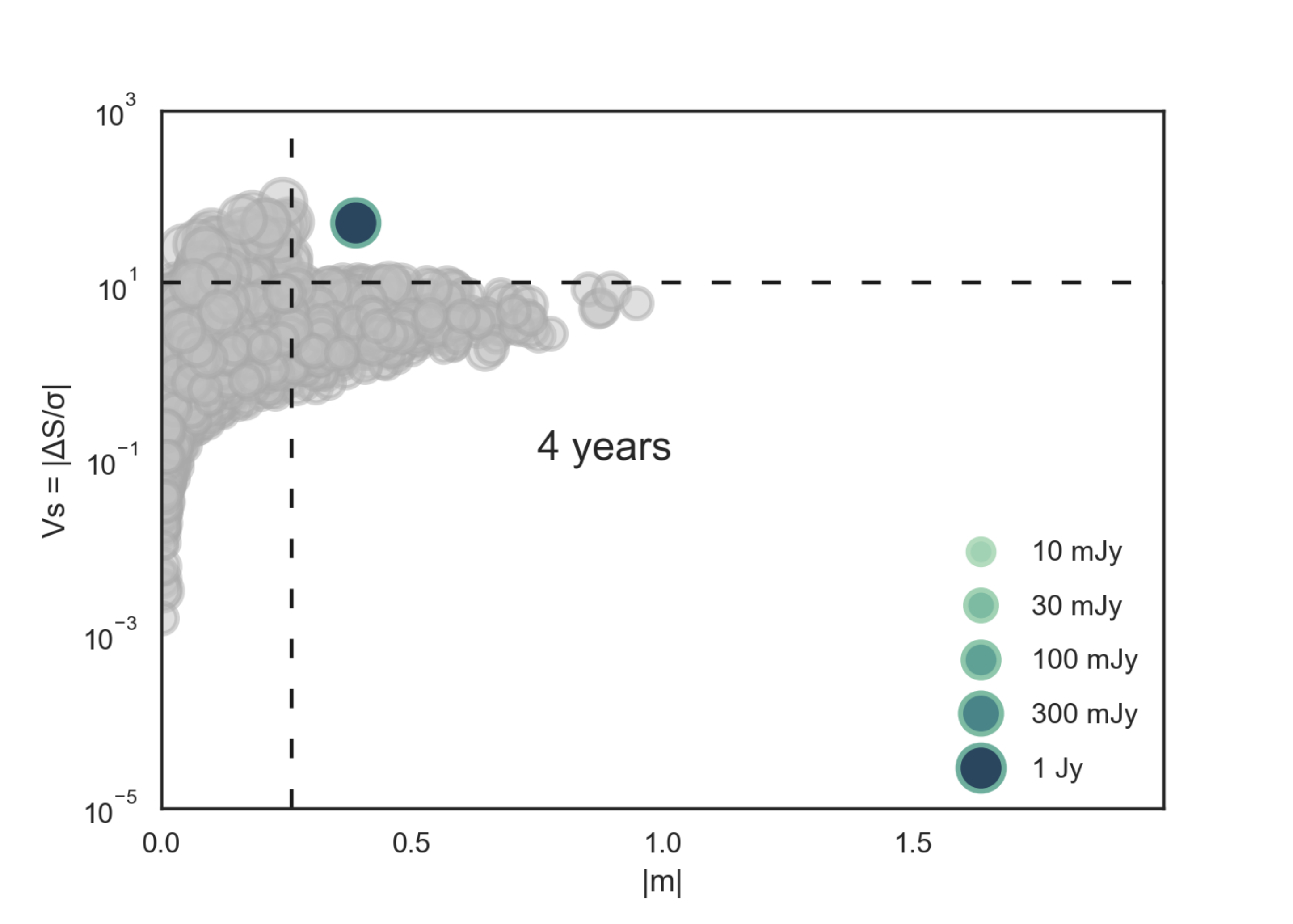}
\caption{The variability statistic, Vs, as a function of modulation index, m, for all timescales probed in this work: $<$4 hours, $<$1 day, $<$1 month and $<$4 years. The dashed lines correspond to final selection criteria i.e. limits on m and Vs. The green-to-blue circles are sources which are finally shortlisted as variables after visual inspection. The size of the circle denotes the mean flux density of the source in two epochs. We find 18, 2 and 12 variables on timescales of 4 hours, 1 day, and 4 years.}
\label{fig:real_var}
\end{figure*}

Following the application of the constraints mentioned above to our 5$\sigma$ PyBDSF catalogs (for each individual image mosaic described above), we used \texttt{TOPCAT} (Tool for OPerations on Catalogues And Tables, v4.6-1; \citealt{topcat}) to perform a two-epoch comparative study at every timescale. Given the synthesized beam of GMRT at 150 MHz, 19$\arcsec$ $\times$ 15$\arcsec$, we used  a search radius of $\sqrt{\texttt{BMAJ}\times\texttt{BMIN}}/2= 9\arcsec$ to find the counterparts between any two epochs. The following `two-epoch' comparisons were successfully performed under the aforementioned conditions:
\begin{itemize}
    \item 4 yr timescale: 2132 two-epoch comparisons (2132 unique sources were matched) between our combined survey data and TGSS-ADR
    \item 1 month timescale: 4686 two-epoch comparisons (4686 unique sources matched) between E1 and E2
    \item 1 day timescale: 6987 two-epoch comparisons (among which 4389 unique sources were matched) for E1a vs. E1b and E2a vs. E2b.
    \item 4 hour timescale: 7134 two-epoch comparisons (among which 6689 unique sources were matched) for E1a scan1 vs. scan2, E2a scan1 vs. scan2, and E2b scan1 vs. scan2.
\end{itemize}

For every source catalog comparison made, we applied a suitable correction factor to ensure that the ratio of the source flux densities between the two epochs (S1/S2) is unity. The median of S1/S2 was taken to be the correction factor and applied to (divided out from) source flux densities and the associated uncertainties in the (fiducial) first comparison epoch (S1). The correction factors ranged between 0.85 (4 hr timescale) and 0.98 (4 yr timescale). 
We then used the corrected source flux densities with the corrected uncertainties to calculate two statistical measures, the variability statistic ($V_s$) and the modulation index ($m$), to distinguish between true variables and false positives. Following \cite{mooley2016}, we compared the flux densities of a source between two different epochs using the $V_s = (S_1 - S_2)/\sqrt{{\sigma_1}^2+{\sigma_2}^2}$ = $\Delta S/\sigma$. The null hypothesis is that the sources are selected from the same distribution and are hence non-variable. Under this hypothesis, $V_s$ follows a Student-t distribution. However, in our case we find that the distribution is Gaussian (see Figure \ref{fig:Vs_hist1}). This may be explained by ionospheric effects in the low-frequency sky, other systematic effects in the amplitude calibration, cleaning artifacts etc.
Nevertheless, we are able to fit Gaussian functions to the $V_s$ distributions, for the four timescales probed, and we consider a source as a true variable if it has $V_s$ lie beyond 4$\sigma$ in the distribution \cite[see][]{mooley2016}. Our criterion for selecting a true variable source is therefore: 

\begin{align}
V_s =\left|\frac{\Delta S}{\sigma}\right|> 4 \times {\tt std}
\end{align}

where ${\tt std}$ is the standard deviation of the $V_s$ distribution (see Figure~\ref{fig:Vs_hist1}). Modulation index, $m$, is a measure of variability defined as difference of flux densities of a source between two epochs divided by the mean of the two flux densities, $\overline{S}$
\begin{align}
m =\frac{\Delta S}{\overline{S}} = 2\times\frac{S_1-S_2}{S_1+S_2}
\end{align}

\begin{figure*}
\includegraphics[width=7in]{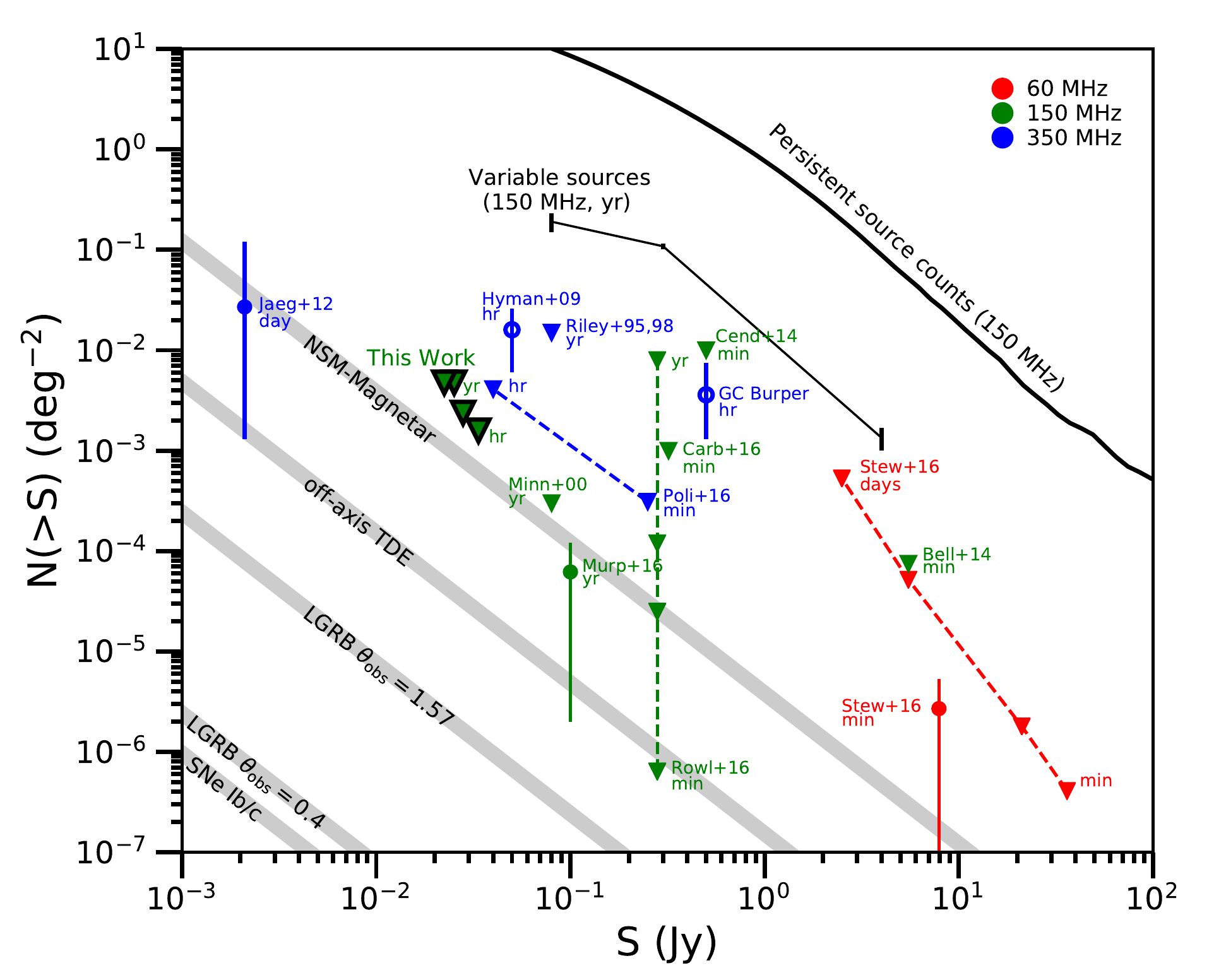}
\caption{The log(N)--log(S) phase space of low-frequency radio transients. The 2$\sigma$ upper limits to the transient rates from previous radio surveys (see the compilation at http://www.tauceti.caltech.edu/kunal/radio-transient-surveys/index.html) are shown as triangles. Rates from the same survey are joined by dashed lines. The rates derived from radio transient detections are shown as 2$\sigma$ errorbars. The extragalactic transient rates, at 150 MHz, from \citet{metzger2015} are shown with thick gray lines. The symbols are color-coded according to observing frequency. The source counts for persistent \citep[from the TGSS-ADR; ][]{intema2017} and variable sources \citep[$m\gtrsim0.1$ at 150 MHz, based on][]{mcgilchrist1990,riley1993,minns2000,bell2019} are shown with black lines. Timescale corresponding to each transient detection or upper limit is denoted as min (minute), hr (hour), day (day), mo (month) or yr (year). References: \citet{bell2014, carbone2016, cendes2014, riley1995, riley1998, polisensky2016, rowlinson2016} (other references are cited in the text). Upper limits from \citet{feng2017}, at 182 MHz and on timescales between minutes and months, lie in the region similar to the \citet{polisensky2016} limits and are not shown on this plot. Transient rate upper limits from our survey, on timescales of 4 hr, 1 day, 1 month and 4 years, are shown as thick green triangles.}
\label{fig:logN-logS}
\end{figure*}

Given the uncertainties in flux calibration, ionospheric effects and the like, we consider a source as a true variable only if the fractional variability is more than or equal to $30\%$ \citep[i.e. a modulation index of $|m|>0.26$; see also][]{mooley2016}. 

We shortlisted the variable candidates using the above criteria. Then we visually inspected the image cutouts (from our survey as well as archival data from NVSS and FIRST) of these candidates and removed the potentially resolved sources.
We thus found $1$ variable for the $4$ year timescale, no variables for the $1$ month timescale, $1$ variable for the $1$ day timescale and $6$ variables for the $4$ hour timescale.  
These variables are shown in Figure~\ref{fig:real_var} (variability statistic against modulation index plots for each timescale probed) and their details are tabulated in Table~\ref{tab:vartransum}. 
The typical modulation index is 0.3--0.4.
Identification of the variable sources and estimation of the variability fraction of the 150 MHz sky is done in \S\ref{sec:summary}. 
\section{TRANSIENT SEARCH}\label{sec:trans_search}

\noindent For our transient search, we chose a higher detection threshold than the 5$\sigma$ used for the variability search. Considering an average, $\sim$18 arcsec, synthesized beam for our survey, and searching effectively across $\sim$4200 sq deg (300 sq deg survey area $\times$ 14 observations searched), implies 50 Million synthesized beams searched. Hence, in order to keep the number of false positives, due to noise, down to $<$1, we chose a 7$\sigma$ source detection threshold for transient search \citep[following the recommendation of][]{frail2012}.
\vspace{1mm}\\
\indent We used the same point-source constraints defined above, for the variables case, to perform the transient search. The cumulative number of sources in our resulting point-source catalogs is 68,964 sources. 
We compared the source catalogs as in the above case of variables, probing timescales of 4 yr, 1 month, 1 day and 4 hours. For each single-epoch catalog pair compared (using \texttt{TOPCAT}), we searched for those sources present in one epoch and absent in the other.
For the resulting transient candidates, we further verified their absence in the combined deep mosaics from our survey, and from archival images from the TGSS, NVSS and FIRST surveys. All of these candidates were SNR$<$15 and were either imaging artifacts (due to the presence of nearby bright sources) or appeared to be resolved sources in the archival radio images. 
We thus find no evidence for any transient sources in our data. 






\section{SUMMARY \& DISCUSSION}\label{sec:summary}

\noindent With the aim of probing deeper into the phase space of transients in the low-frequency radio sky, we observed the SDSS Stripe 82 region at 150 MHz at multiple epochs with the GMRT. Our survey region spans 300 sq. deg (uniformity of RMS noise shown in Figure~\ref{fig:rms_cum}) and the observations are tabulated in Table \ref{tab:gmrt_log}. Using our observations in addition to the archival data from the TGSS-ADR, we were able to perform ``two-epoch" comparisons, to find transients and variables, on four different timescales: 4 hours, 1 day, 1 month and 4 years.
Using 5$\sigma$ source catalogs for each timescale, we generated catalogs of point-like sources using a set of constraints, as described in Section \ref{sec:var_search}. 

We found 6, 1, 0 and 1 sources satisfying our variability criteria (significance greater than 4$\sigma$ and fractional variability larger than 30\%; see \S\ref{sec:var_search}) on timescales of 4 hours, 1 day, 1 month and 4 years respectively. We note that the results for the 4 hour timescale are most uncertain due to modest UV coverage and larger flux calibration uncertainty. This is also the timescale for which we found the largest number of false positives (imaging artifacts), compared to our analysis for other timescales. Hence, the number of true variables on the 4 hour timescale is likely to be far less than 6.
\vspace{1mm}\\
\indent Table \ref{tab:vartransum} lists the variable sources that we found, along with their fluxes from the TGSS-ADR, NVSS and FIRST catalogs, the spectral indices with respect to the NVSS source catalog, and the magnitudes and spectroscopic redshifts of their optical counterparts. We also performed source identification (noted in Table~\ref{tab:vartransum}) based on published optical spectra or WISE colors. We find that all the variable sources are AGN. The spectral indices calculated using the flux density in the NVSS survey are consistent with the typical AGN spectral index of $-0.8$ with the exception of J012528+000505, found on the 4 year timescale, and J022609+012929, found on the 4 hour timescale, which have a flat or inverted radio spectra. Comparison of the 150 MHz flux density of J012528+000505 with recently-published 1.4 GHz flux density ($\sim$800 mJy; \citet{heywood2016}) suggests that the source is consistent with being flat-spectrum, and its 1.4 GHz flux density has decreased by a factor of two with respect to the FIRST and NVSS surveys (observed in the 1990s).



\subsection{Variability of the 150 MHz sky}
\noindent We calculate the fraction of persistent sources that are variable as following: On a timescale of 4 hours, we found 6 significant variables out of a total of 7134 independent ``two-epoch" comparisons (see \S\ref{sec:var_search}). This implies that 0.08\% of the persistent sources are variable, having a fractional variability $\geq$30\%. Due to the UV coverage and flux calibration issues noted above for the 4 hour timescale, we consider this fraction as an upper limit. A single variable source was found in each of the 1 day and 4 year timescales, among a total of 6987 (0.01\% of the persistent sources) and 2132 (0.05\% of the persistent sources) ``two-epoch" comparisons, respectively. No variables were found on the 1 month timescale (among 4686 "two-epoch" comparisons), and if we assume three sources as the 2$\sigma$ upper limit \citep{gehrels1986}, then we get the variability fraction as $<$0.06\% of the persistent sources.


Variability in these sources, listed in Table~\ref{tab:vartransum}, is most likely extrinsic rather than being intrinsic to the sources themselves\footnote{Incoherent emission sets a limit on the brightness temperature, as we discuss below. We do not attribute variability of our sources (all of which are AGN) due to coherent emission since this would require invoking new physics in AGN, which we believe is unlikely.}. One of the suspects could be the ionosphere, but the SPAM pipeline (see \S\ref{sec:obs_proc}) is expected to minimize this factor. 
Interstellar scintillation, on the other hand, is expected to be the dominant factor. Brightness temperature constraints ($T_b\lesssim10^{12}$ K for synchrotron emission; \citet{Kellermann-Pauliny-Toth1969,readhead1994}) place strong limits on the source size of the radio emitting region. Assuming that the source size is comparable to the light travel time c$\tau$, the variability in flux density at 150 MHz is constrained as follows, unless relativistic beaming is involved. 

\begin{equation}
\Delta S \lesssim 0.03\,{\rm mJy} \, (\tau/1\,{\rm yr})^2 (D_A/1.5\,{\rm Gpc})^{-2}
\end{equation}

where $\tau$ is the variability timescale, and $D_A$ is the angular diameter distance. Therefore, any intrinsic component to the variability will be limited to sub-mJy flux densities. None of the variable sources (having optical counterparts) show any evidence of blazar activity in their optical spectra, and therefore we do not expect relativistic beaming.
We thus find extrinsic variability (refractive interstellar scintillation or RISS; consistent with \citealt{rickett1986}) to be the most probable explanation of the flux density changes seen in our sources.

These results are also consistent with previous variability surveys.
For example, \cite{mcgilchrist1990}, \cite{riley1993} and \cite{minns2000} carried out observations of several extragalactic fields with the Rile telescope at 150 MHz, and found 2/811 sources, 21/1050 and 207/6000 sources brighter than $\sim$100 mJy, respectively, to be variable at the $\gtrsim$10\% level on timescales of $\geq$1 yr.
\cite{riley1993} noted enhanced variability in flat-spectrum sources and in steep spectrum sources whose spectra turn over at about 400 MHz.
A similar conclusion was derived by \cite{bell2019}, who recently studied the variability of 944 sources brighter than 4 Jy at 154 MHz with the MWA.
They found 15 sources (1.6\% of the sources monitored) to be variable on a timescale longer than 2.8 years, and noted enhanced variability in sources having peaked spectral energy distributions.
All these studies have attributed the source variability to RISS.
In our sample of variable source, we find 1--2 sources are flat spectrum, while the others are steep spectrum (we cannot exclude the possibility of the latter having spectra peaking at $\sim$100 MHz.)
We mark the variable source counts\footnote{These denote sources varying beyond the $\gtrsim$10\% level. Source counts from our search are much lower, since we considered sources varying only beyond 30\%.} from \cite{mcgilchrist1990}, \cite{riley1993}, \cite{minns2000} and \cite{bell2019} in Figure~\ref{fig:logN-logS}.


The variability of the low-frequency radio sky is substantially lower than that of the GHz sky. A number of studies of the dynamic GHz sky \citep[e.g.][]{carilli2001,thyagarajan2011,bannister2011,croft2011,mooley2013,williams2013,bell2015,mooley2016} have shown that $\sim$1\% of the persistent sources at frequencies of 1--few GHz are variable beyond the $\sim$30\% level, on timescales ranging from days to years. At 150 MHz, the fraction of variables among persistent sources is less by a factor of 10 or more. 

We have attributed the variability of our sources to extrinsic factors, likely RISS.
It is possible that interplanetary scintillation (IPS) may be playing a role, since the Stripe 82 region lies along the ecliptic.
In their study of IPS at 162 MHz, \cite{morgan2018} find modulation indices of $\gtrsim$0.5 for radio sources lying along or in the vicinity of the ecliptic, and m$\lesssim$0.25 for sources lying away from the ecliptic.
Indeed some of the variable sources on 4 hour timescale may also be due to IPS, although the flux scale for this timescale is most uncertain.
Future surveys carried out with the LOFAR, the MWA and the SKA-low will find significant variability resulting from IPS.


\subsection{Transient rates at low frequencies}

\noindent We now calculate the upper limits to the transient rate from our survey. 
Using Poissonian statistics, we take the 2$\sigma$ upper limit to the number of transients as 3. Since we have carried out 6, 4, 2 and 2 two-epoch comparisons on timescales of 4 hours, 1 day, 1 month and 4 years respectively, we calculate the upper limits\footnote{This is calculated as 3/(Area $\times$ epochs), where we take the survey area to be 315 deg$^2$.} as $1.6\times10^{-3}$ deg$^{-2}$, $2.4\times10^{-3}$ deg$^{-2}$, $4.8\times10^{-3}$ deg$^{-2}$ and $4.8\times10^{-3}$ deg$^{-2}$ respectively (these are the instantaneous snapshot rates). The quoted upper limits to the transient rate are for 7$\sigma$ flux density thresholds, i.e. 28 mJy, 34 mJy, 22 mJy and 25 mJy respectively. 

In Figure~\ref{fig:logN-logS} we show the log N($>$S)-log S phase space of the dynamic low-frequency radio sky (S is the flux density and N is the number of radio sources). Persistent source counts from the TGSS-ADR are shown as a thick black line.
The transient rate upper limits (including those from our survey) and detections from past blind searches below 400 MHz are plotted as triangles and errorbars.
For reference, the rates of extragalactic transients considered by \cite{metzger2015}, assumed to follow a Euclidean $N(>S)\propto S^{-1.5}$ distribution, are plotted as grey shaded areas. 
The symbols are color coded to represent observing frequency.
Searches that were primarily extragalactic are shown with filled symbols and those that were primarily Galactic (mainly towards the Galactic Center) are shown with unfilled symbols.

\subsection{Investigation the radio transient phase space and recommendations for future low-frequency transient surveys}
\noindent We make the following observations from Figure~\ref{fig:logN-logS} and make recommendations for maximizing the yield of transients at low radio frequencies.

Firstly, the rate of Galactic Center transients, such as the ``burper" \citep{hyman2005,kulkarni2005} and the X-ray binary found by \cite{hyman2009}, is significantly larger than the rate of extragalactic transients.
The rate is higher by a factor of $\gtrsim$10.
This suggests that low-frequency radio surveys of the Galactic Center, Galactic bulge or the Galactic plane will be lucrative.

Secondly, although we have sampled a competitive part of the phase space (where the population(s) uncovered by \cite{jaeger2012}, \cite{murphy2017} and \cite{stewart2016} reside(s), assuming $N\propto S^{-1.5}$ distribution) with our medium-deep medium-wide GMRT Stripe 82 survey, we have not recovered any transients\footnote{Our transient search on the $<$4 hour timescale is capable of finding transients similar to the one found by \cite{stewart2016} (which had a timescale of a few minutes timescale), since each of our observations, that were compared, was 10--20 minutes long. This of course assumes that the emission is broadband and the spectral index of the transient between 60 MHz and 150 MHz is not very steep.}.
This suggests that a multi-epoch survey covering $\gtrsim$1000 deg$^2$ may be required to find any transient, in extragalactic fields, at the $\sim$10 mJy sensitivity level.

Our survey together with the transient rate upper limits on minutes/hour timescales from \cite{rowlinson2016} (both surveys carried out at around 150 MHz) suggest that the transient class detected by \cite{stewart2016} (at 60 MHz; assuming that the source is astrophysical) either 1) does not follow a Euclidean distribution or 2) has a steep spectrum or narrowband emission. 
Otherwise, we would have expected to find at least a few such transients in the 150 MHz surveys.
We define null probability as the probability of not detecting any transients (of a particular class) in our survey. 
Assuming Poisson statistics and Euclidean distribution, we derive a null probability for \citeauthor{stewart2016}-like transients of $\ll$1\%.
It is possible that such events may be caused by variability (intrinsic or extrinsic) of compact Galactic sources (for which we speculate that the source counts are flat ($N(>S) \propto S^{-1}$ or $\propto S^{-0.5}$) because the source density falls off substantially beyond a distance of a few kpc. 
In this case, we expect the rate of such events to be high close to or within the Galactic plane, and this possibility can be explored with Galactic plane transient surveys at low radio frequencies.
If we attribute the absence of these transients in our survey and in \cite{rowlinson2016} purely to steep spectral index (while assuming $N\propto S^{-1.5}$), then we calculate the spectral index constraint to be $\alpha\lesssim-4$.

The implied rate of the transients like the one found by \cite{jaeger2012} is N($>$1 mJy)=0.1 deg$^{-2}$.
In the GHz sky, the only transient class known to have such a high rate is active stars and binaries \citep[e.g.][]{mooley2016}.
Hence, we advocate that the \citeauthor{jaeger2012} transient is a stellar flare, otherwise a different emission mechanism needs to be invoked.
A stellar flare interpretation is also consistent with the \cite{murphy2017} transient, whose implied snapshot rate per deg$^{2}$ is similar to the \citeauthor{jaeger2012} transient, and was found at low Galactic latitude. This is in line with the M dwarf counterpart/candidate ($d\sim1.5$ kpc in Gaia; \citet{Gaia2018}) proposed by \citeauthor{murphy2017}.
The null probabilities of finding transients, like the ones uncovered by \citeauthor{jaeger2012} and \citeauthor{murphy2017}, in our survey are approximately 2\% and 40\% respectively.

As discussed earlier in this section, the transient upper limits from our GMRT survey advocate Galactic searches or very widefield extragalactic searches. 
We therefore provide recommendations for maximizing transient discovery using existing low-frequency radio interferometers.
Considering their modest fields of view ($\ll$100 deg$^2$), widefield surveys will be expensive to execute with telescopes such as the GMRT, LOFAR, especially given the computing time/cost for data processing.
Hence, we recommend surveys of the Galactic plane or Galactic Center for these telescopes. 
The geographical location and the recent upgrade of the GMRT makes the observatory uniquely situated to carry out sensitive surveys of the Galactic Center with arcsecond localization capability.
Although extragalactic transients will be challenging to find with such telescopes, searching for the radio afterglows of neutron star mergers (detected as gravitational wave sources) over tens of square degree localization regions may be worthwhile, especially since reference images can now be provided by the LoTSS \citep{shimwell2019} and TGSS-ADR \citep{intema2017}. 

Widefield surveys with the MWA or with the VLA (VCSS, currently being undertaken alongside the VLASS) may be useful for finding old, optically thin extragalactic transients (the transient found by \citealt{murphy2017} may be one such event) and constraining the rates of such transients. All-sky imagers like the LWA1 and OVRO-LWA will be excellent for finding big samples of transients similar to \citealt{obenberger2014}, thus identifying these transients with a known class of objects, as well as for detecting coherent emission from Galactic sources and the mergers of neutrons stars. Eventually, SKA-low will be able to routinely survey the low-frequency sky and provide a complete census of the dynamic Galactic and extragalactic sky.

\begin{table*}
\footnotesize
\centering
\caption{Summary of variables sources.}
\label{tab:vartransum}
\begin{tabular}{lccccccccccccc}
\hline\hline 
Name & RA    & DEC   & S1    & S2    & m & Vs  & S$_{\rm TGSS}$ & S$_{\rm NVSS}$ & S$_{\rm FIRST}$ & $\alpha_{0.15}^{1.4}$   & Ident. & r & spec-z\\
(G1STS J...) & (deg) & (deg) & (mJy) & (mJy) &	 &	  &	 (mJy)  &	  (mJy)      &	  (mJy)       & 	     &  & (mag)   \\
\hline
\multicolumn{14}{c}{Timescale $<$ 4 years}\\ %
\hline
012528+000505 & 21.3699 & -0.0990 & 493 $\pm$ 3 & 731 $\pm$ 3 & 0.39 & -52.0  & 731 & 1540 & 1401 & 0.41  & QSO & 16.5 & 1.08\\  
\hline
\multicolumn{14}{c}{Timescale $<$ 1 month}\\
\hline
\multicolumn{14}{c}{None}\\

\hline
\multicolumn{14}{c}{Timescale $<$ 1 day}\\
\hline

004608+000505 & 11.5355 & 0.0935 & 478 $\pm$ 5 & 627 $\pm$ 8 & 0.27 & 15.6 & 519 & 96 & 87 &-0.78 & QSO & 20.3 & 1.44\\ 
\hline
\multicolumn{14}{c}{Timescale $<$ 4 hours$^*$}\\
\hline
022109+002525 & 35.2893 & -0.4296 & 343$\pm$5 & 445$\pm$3 & 0.26 & 15.8  & 331 & 335 & 313 &  -0.07 & AGN & 20.5 & 0.48  \\


022609+012929 & 36.5402 & 1.4906 & 1111 $\pm$ 12 & 776 $\pm$ 15 & 0.36 & 16.9  & 1247 & 363 & 340 & -0.43 & QSO & 18.5 & 1.37  \\

013227+002828 & 23.1165 & -0.4766 & 293 $\pm$ 5 & 153 $\pm$ 7 & 0.63 & 15.2  & 316 & 66 & 50 & -0.54 & AGN & 24.6 & - \\

012205+000808 & 20.5248 & 0.1497 & 1073 $\pm$ 6 & 820 $\pm$ 9 & 0.27 & 22.7  & 1309 &172 &156 &-0.76&  AGN & - & -\\

225224+012626 & 343.1039 & 1.4394 & 225 $\pm$ 6 & 390 $\pm$ 6 & 0.54 & -19.5  & 382 & 52 & 49 & -0.79 & AGN & - & - \\

223908+012020 & 339.7868 & 1.3410 & 185 $\pm$ 5 & 294 $\pm$ 4 & 0.45 & -15.9 & 237 & 51 & 44 &-0.68 & AGN & 21.3 & 0.53 \\
\hline
\multicolumn{10}{l}{$^*$The flux scale is most uncertain for this timescale. Many of these variable candidates may be false positives. See \S\ref{sec:summary}.}\\
\end{tabular}
\end{table*}

\quad \newline \newline
{\it AH acknowledges support from the 2018 NRAO summer research program, where the majority of the analysis was done. KPM is a Jansky Fellow of the NRAO. Thanks to Nimisha Kantharia, Preshanth Jagannathan and Gregg Hallinan, who provided help with the GMRT proposal and observing. We thank the staff of the GMRT who have made these observations possible. The GMRT is run by the National Centre for Radio Astrophysics of the Tata Institute of Fundamental Research. We thank the anonymous referee for helpful comments on the manuscript.}

\end{document}